\documentclass[twocolumn,preprintnumbers,amsmath,amssymb]{revtex4}
\usepackage{graphicx}
\usepackage{dcolumn}
\usepackage{bm}
\usepackage{epsfig} 
\begin{document}


 


\title{Characterization of the material response in the granular ratcheting}

\author{ R. Garc\'{\i}a-Rojo }
\homepage{http:\\www.icp.uni-stuttgart.de/~rojo}
\email{rojo@icp.uni-stuttgart.de}
\affiliation{ICP, University of Stuttgart, \\
Pfaffenwaldring 27, 70569 Stuttgart, Germany\\
}

\author{F. Alonso-Marroqu\'{\i}n}
\email{fernando@icp.uni-stuttgart.de}
\affiliation{ICP, University of Stuttgart, \\
Pfaffenwaldring 27, 70569 Stuttgart, Germany\\
}
\author{H. J. Herrmann}       \email{hans@icp.uni-stuttgart.de}
\affiliation{ICP, University of Stuttgart, \\
Pfaffenwaldring 27, 70569 Stuttgart, Germany\\
}

\date{\today}

\begin{abstract}
The existence of a very special ratcheting regime has recently been reported
in a granular packing subjected to cyclic loading~\cite{alonso04}. 
In this state, the system accumulates a small 
permanent deformation after each cycle. After
a short transient regime, the value of this permanent strain accumulation
becomes independent on the number of cycles. We show that a characterization of the material response in this peculiar state is possible in terms of three simple macroscopic variables. They are defined that, they can be easily measured both in the experiments and in the simulations. We have carried out a thorough investigation of the micro- and macro-mechanical factors affecting these variables, by means of Molecular Dynamics simulations of a polydisperse disk packing, as a simple model system for granular material. Biaxial test boundary conditions with a periodically cycling load were implemented. The effect on the plastic response of the confining pressure, the deviatoric stress and the number of cycles has been investigated. The stiffness of the contacts and
friction has been shown to play an important role in the overall response of the system. Specially elucidating is the influence of the particular hysteretical behavior in the stress-strain space on the accumulation of permanent strain and the energy dissipation.

\end{abstract}
\maketitle

\section{Introduction}
\label{introduction}

Brownian motors, quantum ratchets or molecular pumps, all these machines
operate under the same principle: The chaos of the micro-world cannot be
avoided, but one can take advantage of it \cite{reimann02}. 
Nanoscale  ratchet devices have been designed with the surprising
property that they can  extract work from the noise of thermal
and quantum fluctuations~\cite{zapata96}. 
Ratcheting is the mechanism behind  molecular motors, which can use the 
chaotic Brownian motion to turn directionless energy into directed motion
\cite{howard97}. These lilliputian motors seem to be 
responsible for many biological  process, such as mechanical  
transport~\cite{svoboda93} or muscle contraction~\cite{kitamura99}. 
Apart from these  fascinating 
machines, the ratchet effect has been used to describe economical or  
sociological processes where the intrinsic asymmetry in 
the system allow  to rectify an unbiased input \cite{dybvig95}. 
 A ratchet-like effect is also the major cause of material deterioration due to cyclic stress loading, thermal or mechanical 
fluctuations \cite{lekarp00, royer04, sou00}.
Asymmetries in foundations can produce tilting  and eventual collapse 
of any structure due to ratcheting \cite{england95}. 
The tower of Pisa is a well documented case, where
the tilt was observed from its construction in 1173~\cite{burland94}. Pavement design is another important field in which graded soils are used 
as supportive roadbed \cite{lekarp98,lekarp00,sharp84,werkmeister01}. 
The excitations that traffic imposes on the sub-layer 
produce deformations in the granular material. These deformations are 
transmitted to the upper layers of the pavements, causing its degradation or 
even its breakage. Cyclic loading tests are extensively used in the 
investigation of the plastic response of unbound granular matter \cite{lekarp98} . In these experiments, the material is subjected to a certain cyclic stress condition mimicking traffic. From a practical point of view, the main question is whether the material accumulates plastic deformation in each cycle, or 
whether it adapts to the excitation reaching a shakedown state. Only materials in which the excitations {\it shake down} should be consequently used in pavement design.

The use of simple models of granular materials allows the numerical solution 
of the dynamics. Discrete Element Methods  (DEM) such as
Molecular Dynamics (MD) \cite{cundall89,luding04,oda97} and 
Contact Dynamics (CD) \cite{radjai96,moreau94}  have been in fact often successfully applied to the investigation of the elasto-plastic behavior of granular matter. Specially interesting from the physical point of view, is how the contact modelization affects the overall response \cite{coste04,farkas02}.  Recent MD results have shown the key role that sliding plays on the plastic deformation of a granular packing subjected to cyclic loading, and the existence of a range of values of the excitations for which a simple visco-elastic model of disks subjected to cyclic loading attains shakedown \cite{alonso04, garcia-rojo04}. Beyond the shakedown limit, two other  possible responses have also been identified; 
For very high loads, the material accumulates deformations at a relatively 
high constant rate, leading to an incremental collapse of the structure. 
For moderate loading intensities, the system undergoes an adaptation 
process in which the accumulation of deformation gradually decreases to 
a very low constant value. This 
post-compaction is associated to a relaxation of the dissipated energy per 
cycle, that progressively decreases to a constant value dependent on the 
imposed loading. In this final stage, there is a small but persistent 
accumulation of permanent strain, associated to a periodic behavior of the 
sliding contacts \cite{alonso04}, which is called ratcheting regime.

Due to the non-lineality and the irreversibility of the behavior, cyclic 
loading is a rather complicated problem from the theoretical point of view. 
Elasto-plastic and hypoplastic theories can account for the change in the 
incremental stiffness during loading and unloading phases, only if basic 
modifications are undertaken \cite{kolymbas99,tatsuoka03}. In the case of 
elastoplasticity, the overall plastic behavior in the loading-unloading 
is obtained as the result of a combination of several {\it yield surfaces}
\cite{mroz81}. 
In the hypoplastic theory the {\it inter-granular strain} is introduced to take into account the dependence of the response on the deformation history~\cite{niemunis96}. Interestingly, a point of convergence of both theories has been established by the {\it bounding surface elasto-plasticity} \cite{dafalias86a}. This theory introduces a tiny elastic nucleus changing with the deformation,  and describe the hysteresis by means of internal variables taking into account the evolution of the microstructure. The characterization of such internal variables has been traditionally done using structure tensors, measuring the fabric properties of the contact network \cite{thornton86}. There is numerical evidence that a single fabric tensor, measuring the anisotropy of the contact network can be used to characterize the resilient response \cite{cowin85}. But the description of the plastic deformation requires to take into account the inherent decomposition of the contact network in sliding and non-sliding contacts \cite{alonso05c}. The role of kinematical modes such as sliding and rolling  has been also investigated to some extent for monotonic deformation, but not for cyclic loading \cite{astrom00,cundall89,latzel03}.

The final aim of this paper is the characterization of ratcheting response of a granular packing under cyclic loading. For this purpose, three macroscopic variables will be introduced. A simple DEM model will then be used to investigate the dependence of the material response on different macroscopic and microscopic variables. From this investigation, we have found that our simple model is able to reproduce several behaviors observed in the experience, and microscopically justifies the use of popular empirical laws, like the $k- \theta$ model. The main parameters of our model and the details of the MD simulations are 
presented in  Section~\ref{model}.  
The ratcheting regime resulting in the biaxial test  is described  in Sect. 
\ref{ratcheting}.  In Sect.~\ref{material} we decompose the 
strain response in its  permanent and resilient components. We continue with an analysis of hysteresis in the plastic response, establishing in Sect.~\ref{hysteresis}, a direct relation between the particular shape of the stress-strain cycle and the dissipated energy per cycle. From this relationship it will be easy to explain the observed dependence of the dissipated energy per cycle on the deviatoric stress. Results on 
the permanent strain and the resilient parameters are presented for the 
different cases studied in Secs. \ref{plastic} \& \ref{resilient}. The approach proposed here is basically empirical.  
The resilient parameters will be therefore conveniently defined in terms of the recoverable deformation, as is usually done by experimentalists
\cite{lekarpb00}. The dependence on the imposed stress is investigated, and
the results are compared to predictions of resilient response models 
\cite{hicks71, allen74, uzan85,  tam88}. The influence of the friction and the stiffness at the contacts, main micro-mechanical parameters of the model, will also be determined.  We finish in Section \ref{discussion} with a discussion of the main conclusions of this work.


\section{model}
\label{model}

In our visco-elastic 2D model, the grains are modeled by soft disks. The 
deformation that two grains suffer during the interaction is reproduced by 
letting the disks overlap. During the overlapping, a certain force $f^c$ is exerted at the contact point. This force can be decomposed in the following parts: 
\begin{equation}
\vec{f}^c=\vec{f}^e+\vec{f}^v,   
\label{eq:contact force}
\end{equation}

\noindent
where $\vec{f}^e$  and  $\vec{f}^v$ are the elastic and viscous 
contribution.  The elastic part of the contact force is also decomposed 
as 

\begin{equation}
\vec{f^e}= f^e_n \hat{n}^c + f^e_t \hat{t}^c. 
\label{eq:elastic force}
\end{equation}

\noindent
The unit normal vector $\hat{n}^c$ points in the direction of the vector
connecting the center of mass of the two disks. The  tangential vector 
$\hat{t}^c$ is perpendicular to $\hat{n}^c$.
The normal elastic force is calculated as 

\begin{equation}
f^e_n= -k_n A/L_c,
\label{eq:normal force}
\end{equation}

\noindent
where $k_n$ is the normal stiffness, $A$ is the overlapping area and 
$L_c$ is a characteristic length of the contact.  
Our choice is $L_c=R_i+R_j$. This normalization is necessary to 
be consistent in the units of force.

The frictional force is calculated using an extension of the method 
proposed by Cundall-Strack \cite{cundall79}. An elastic force   
proportional to the elastic displacement is included at each contact 

\begin{equation}
f^e_t= -k_t \Delta x^e_t,
\label{eq:tangential force}
\end{equation}

\noindent
where $k_t$ is the tangential stiffness.  The elastic displacement 
$\Delta x_t $ is calculated as the time 
integral of the tangential velocity of the contact during the 
time where the elastic condition  $|f^e_t|<\mu f^e_n$ is satisfied. 
The sliding condition is imposed, keeping this force constant when 
$|f^e_t|=\mu f^e_n$. The straightforward calculation of this elastic 
displacement is given by the time integral starting at the beginning 
of the contact:

\begin{equation}
\Delta x^e_t=\int_{0}^{t}v^c_t(t')\Theta(\mu f^e_n-|f^e_t|)dt',
\label{friction} 
\end{equation}

\noindent
where $\Theta$ is the Heaviside step function and $\vec{v}^c_t$ 
denotes the tangential component of the relative velocity $\vec{v}^c$ 
at the contact:

\begin{equation}
\vec{v}^c=\vec{v}_{i}-\vec{v}_{j}+\vec{\omega}_{i}\times\vec{R}_{i}
-\vec{\omega}_{j}\times\vec{R}_{j}.
\end{equation}

\noindent
Here $\vec{v}_i$ is the velocity and  $\vec{\omega}_i$ is the  
angular velocity  of the particles in contact. The branch vector 
$\vec{R}_i$ connects the center of mass of particle $i$ to
the point of application of the contact force. Replacing Eqs.
(\ref{eq:normal force}) and (\ref{eq:tangential force}) into 
(\ref{eq:elastic force}) one obtains:

\begin{equation}
\vec{f^e}= - k_n \frac{A}{L_c} \hat{n}^c - k_t \Delta x^e_t \hat{t}^c. 
\label{eq:elastic force2}
\end{equation}

Damping forces are included in order to allow rapid relaxation 
during the preparation of the sample, and to reduce the acoustic 
waves produced during the loading. These forces are calculated as 

\begin{equation}
\vec{f}^v  = -m(\gamma_n v^c_n \hat{n}^c + \gamma_t v^c_t \hat{t}^c),
\label{eq:viscous force}
\end{equation}

\noindent
being $m=(1/m_i+1/m_j)^{-1}$  the effective mass of the disks
in contact. $\hat{n}^c$ and $\hat{t}^c$ are the normal and tangential 
unit vectors defined before, and $\gamma_n$ and $\gamma_t$ are the 
coefficients of viscosity.  These forces introduce time dependent effects
during the loading. However, these effects can  be arbitrarily reduced by 
increasing the loading time, as  corresponds to the quasi-static approximation.

The interaction of the disks with the walls  is modeled by using a simple 
visco-elastic force:  First, we allow the disks to penetrate the walls. 
Then we include a force

\begin{equation}
  \label{box}
  \vec{f}^b= - \left ( k_n\delta + \gamma_b m_\alpha v^b \right ) \vec{n},
\end{equation}

\noindent
where $\delta$ is the penetration length of the disk, $\vec{n}$ is 
the unit normal vector to the wall, and $v^b$ is the relative 
velocity of the disk with respect to the wall.

The evolution of the position $\vec{x}_i$ and the orientation 
$\varphi_i$ of the particle $i$  is governed by the equations of motion:

\begin{eqnarray}
 m_i\ddot{\vec{x}}_i &=&\sum_{c}\vec{f^c_i}  
+\sum_{b}\vec{f}^b_i, \nonumber\\
I_i\ddot{\varphi}_{i} &=&\sum_{c}\vec{R}^c_i\times\vec{f^c_i}
+\sum_{b}\vec{R}^b_i\times\vec{f}^b_i. 
\label{eq:newton}
\end{eqnarray}

Here $m_i$ and $I_i$ are the mass and moment of inertia of the disk. 
The first sum goes over all those particles in contact with this particle; 
the second one over all the forces given by the walls.
The interparticle contact forces  $\vec{f^c}$ are given 
by replacing Eqs. (\ref{eq:elastic force2}) and (\ref{eq:viscous force}) in 
Eq. (\ref{eq:contact force}).

We use a fifth-order Gear predictor-corrector method for solving 
the equation of motion \cite{allen87a}. This algorithm consists of 
three steps. The first step predicts position and velocity of 
the particles by means of a Taylor expansion. The second step 
calculates the forces as a function of the predicted positions 
and velocities. The third step corrects the positions and
velocities in order to optimize the stability of the algorithm. 
This method is much more efficient than the simple Euler approach 
or the Runge-Kutta method, especially for cyclic loading, where very high 
accuracy is required.

The relevant contact parameters of this model are the normal stiffness at the contacts $k_n$, the ratio of tangential and normal stiffness 
$k_t/k_n$, the normal and tangential damping frequencies and the friction 
coefficient. In the quasi-static approximation, the results are independent 
of the frequency of the cyclic loading. The system is polydisperse, being the radii 
of the grains Gaussian distributed with mean value of $1.0cm$ and variance
of $0.36$.

\section{Onset of granular ratcheting}
\label{ratcheting}

In a biaxial experiment, the sample is subjected to a certain stress state 
characterized by the principal stresses $\sigma_1$ and $\sigma_2$. 
In this case the stress space is therefore a plane, 
since the third component is zero, $\sigma_3 \equiv 0$. 
In our simulations, the system is first  homogeneously compressed with 
$\sigma_1=\sigma_2$. After an equilibrium state under the pressure 
$P_0=\frac{\sigma_1+\sigma_2}{2}=\sigma_1$ has been 
reached, the vertical stress is quasi-statically changed:

\begin{equation}
\sigma_2(t) = P_0 \left[ 1 + \frac{\Delta \sigma}{2} \left( 1- \cos \left( 
\frac{2 \pi t}{t_0} \right) \right) \right],
\end{equation}
where $t$ is the simulation time and $t_0$ is the period of the loading. Note that $\Delta \sigma$, introduced in the last equation, is the maximum deviatoric stress measured in units of $P_0$. In our approximation, it fully characterizes the intensity of the cyclic load imposed on the walls.

Deformation appears in the sample due to the imposed
excitations. The strain is the magnitude that characterizes the 
accumulation of permanent deformation in the sample. Among the different 
practical definitions of strain available
\cite{desai84}, we have chosen here Cauchy's 
definition, which is basically the ratio of the new and the original 
length of the system. Let $L^i_0$ be the original 
length of the sample in the principal direction $i$ ($i={x,y}$). The 
principal component of the strain tensor $\epsilon_{ij}$ on this direction 
will then be:

\begin{equation}
\epsilon_i (t)\equiv \epsilon_{ii}(t) =\frac{L_i (t)-L^i_0}{L^i_0},
\label{eq:defstrain}
\end{equation}
where $L_i$ is the length of the system in the principal direction $i$ at the
moment of the measurement.

Different loading intensities will be exerted on the sample by changing the
value of $\Delta \sigma$. The reaction of the system will be characterized by 
the  deviatoric permanent strain, $\gamma$, that is the 
difference between the strains in the principal directions:

\begin{equation}
\gamma=\epsilon_2-\epsilon_1.
\label{eq:gamma}
\end{equation}

\begin{figure} [t]
\begin{center}
   \epsfig{file=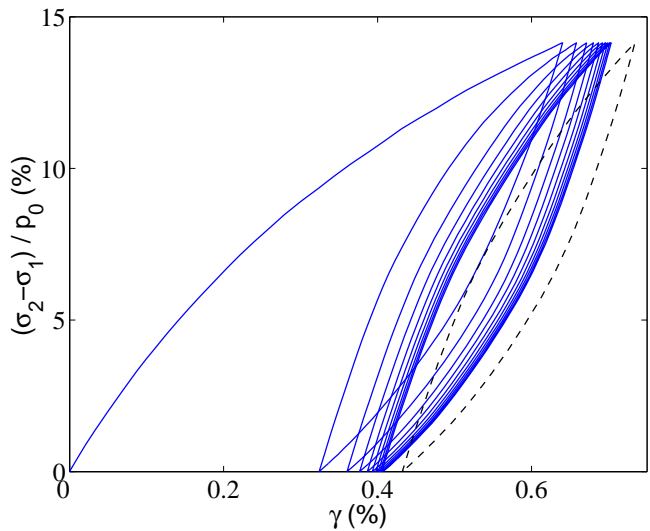,width=1.0\linewidth}
   \caption{Typical stress-strain relation during cyclic loading. In the
long-time behavior the response is given by a limit hysteresis loop. This
is shown by the dashed line ( the loop here corresponds to $N=1000$). In this simulation $\Delta \sigma = 0.14 P_0$ and $P_0=10^{-4} k_n$, where the normal contact stiffness is $k_n= 2 \cdot 10^6 N/m$. The damping constants are defined in terms of the
    characteristic oscillation period $t_s=\sqrt{k_n/\rho \lambda^2}$ (in our case $t_s=0.1414$), where $\rho$ is the density of the grains and $\lambda$ the mean radius of the disks composing the sample. The period of oscillation was taken long enough ($t_0=10^5 t_s$), to be sure that we are in the quasi-static limit).}
   \label{fig:shear}
\end{center}
\end{figure}

\begin{figure} [b]
\begin{center}
    \epsfig{file=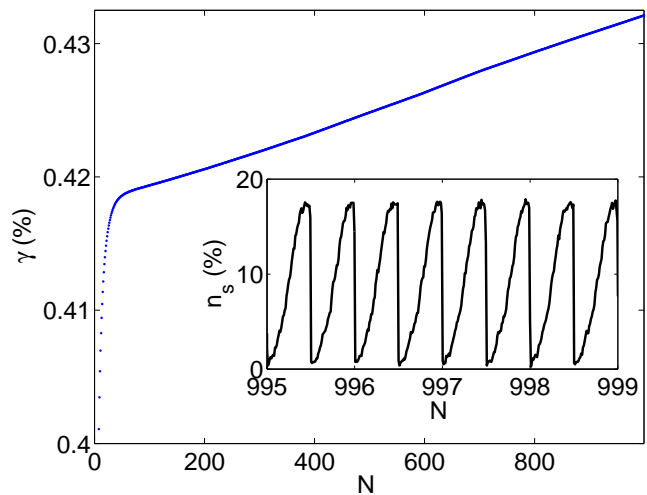,width=1.0\linewidth}
    \caption{Cumulative permanent deformation against the number of 
cycles (N). After the post-compaction regime the system accumulates permanent strain at a constant strain-rate. This is the so-called ratcheting regime, 
which emerges as a result of the periodicity of the sliding contacts.
The inset precisely shows the fraction of the  sliding contact versus time in this state.}
    \label{fig:ratcheting}
\end{center}
\end{figure}

The typical evolution of the permanent strain during the cyclic 
loading is shown in  Fig. \ref{fig:shear}.  The stress-strain  relation 
consists of hysteresis loops. This hysteresis produces an accumulation of 
deviatoric strain with the number of cycles in addition to a progressive 
compaction, which is not shown there.  After some decades of cycles, 
the accumulation of permanent deformation becomes linear, as shown in 
Fig. \ref{fig:ratcheting}. This strain rate remains constant for very large 
number of cycles,  even when the volume ratio is very close to the saturation 
level.

A micro-mechanical explanation of this linear accumulation of strain 
is provided by following the dynamics of the contact network. 
Although most of the contact forces of this network satisfy the  elastic 
condition   $|f_t| < \mu f_n$, the  strong heterogeneities  produce a 
considerable  amount of contacts reaching the sliding condition 
$|f_t| = \mu f_n$ during the compression. After a number of loading
cycles, the contact network reaches a quasi-periodic behavior. In this
regime, a fraction of the contacts reaches almost periodically the sliding 
condition, as shown in the inset of Fig. \ref{fig:ratcheting}. In each load-unload transition there is an abrupt reduction of sliding contacts, which induces the typical discontinuity of the stiffness upon reversal of the loading.  The load-unload asymmetry at each sliding contact makes it to slip the same amount and in the same direction during each loading cycle, leading to an  overall ratcheting response.  

The contact behavior can be observed by embedding two 
points at each particle near to the contact area, and following their translation during each cycle. Their relative displacements are calculated as $\vec s^{i} = \vec s_0 - \vec s_{rb}$, where $\vec s_0$ is the displacement of 
the embedded point $i$ and $\vec s_{rd}$ is the rigid body motion. This latter
is given by the vector connecting the initial to the final position of 
the contact point. Note that $\vec s = 0$ when the two particles move as a 
rigid body.

\begin{figure} [t]
\begin{center}
    \epsfig{file=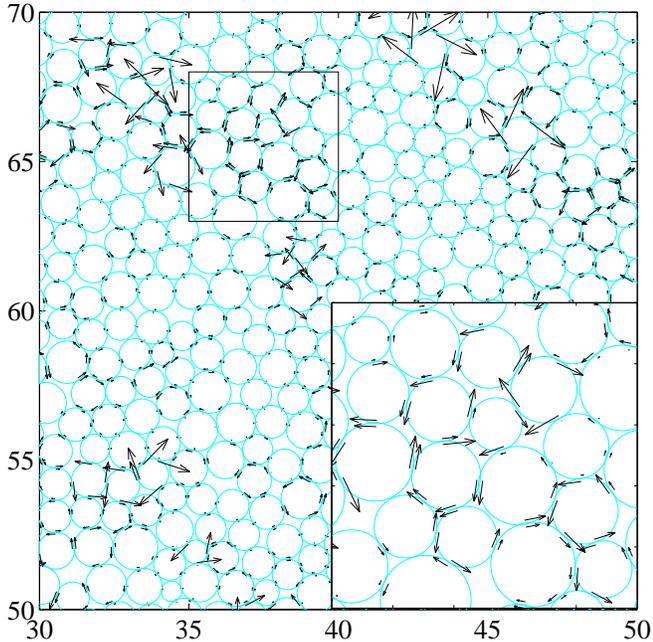,width=1.0\linewidth}
    \caption{Displacement at the contacts during one cycle in the ratcheting. The arrows are proportional to the displacements $s$ of the two material points at the contacts referred to the contact point. More details are found in the text. The figure is a snapshot of the simulation of Figure \ref{fig:shear} for $N=1000$.}
    \label{fig:cosserat}
\end{center}
\end{figure}

Figure \ref{fig:cosserat} shows the displacement at the contacts during cycle $N=1000$. Simulations show that, in this regime, this displacement field is almost constant after each cycle.  There are two deformation modes resembling the  mechanical ratchets: 
(i) At the sliding contacts the  displacement vectors do not agree, so that 
there is a systematic slip during each cycle  which also leads to a
constant  frictional dissipation per cycle.
(ii) At the non-sliding contacts  the displacement vectors are almost the same for the two particles. 

Note from Fig.~\ref{fig:cosserat} that the distribution of this ratchets 
are not uniform, but they are localized in layers resembling shear bands. This kind of strain localization with intense rolling is typical in sheared granular materials \cite{vardoulakis95,astrom00}. Fundamental differences are however observed between the  cyclic loading response and the behavior under monotonic shear: The translation of each particle during the ratcheting regime is given by an almost constant displacement per cycle. On the other hand, the displacement of the particle during monotonic shear is rather chaotic, well described by an anomalous diffusion \cite{radjai02}. 

\begin{figure} [htt]
\begin{center}
    \epsfig{file=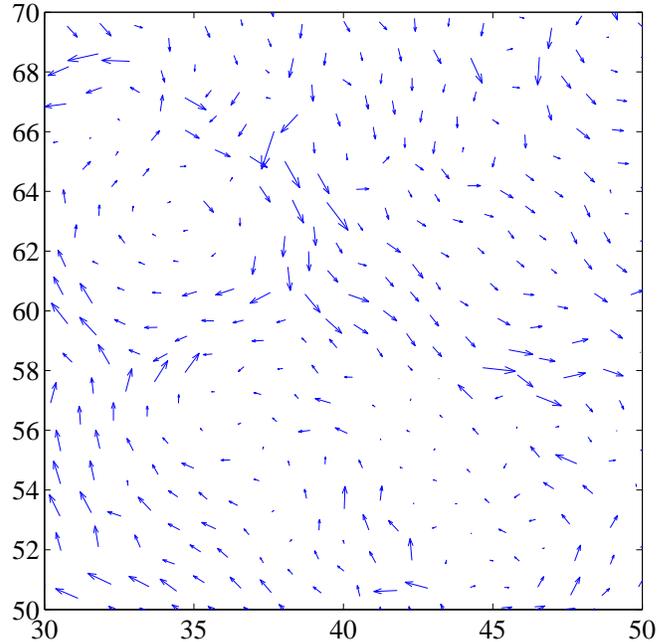,width=1.0\linewidth}
    \caption{Vortex formation as a consequence of the ratcheting of the particles. The arrows are proportional to the displacement of the particle after one cycle in the ratcheting regime. They are plotted at the center of the disks. The cycle is the same as the one shown in Figure \ref{fig:cosserat}.}
    \label{fig:disfield}
\end{center}
\end{figure}

Such systematic translation per cycle of the individual grains in the ratcheting regime has a strong spatial correlation. This is shown in the displacement field of Fig. ~\ref{fig:disfield}. The most salient feature here  is the formation of vorticity cells, where a cluster of  particles rotates as a whole. These vorticities  survive during several hundred of cycles. This is contrary to the  case of the simple  shear, where the vorticities have a very short life-time  ~\cite{radjai02}.   It is interesting to see from Figures ~\ref{fig:cosserat} and ~\ref{fig:disfield} the kinematic phase separation of the grains:  (a) Grains organized in large vorticity cells, and  (b) grains which accommodate the cells to make them more  compatible with the imposed  boundary  conditions. Since such kinematical modes are linked with the a non-vanishing antisymmetric part of the displacement gradient, the strain tensor is not sufficient to provide a  complete description of this convective motion during cyclic loading.  An appropriate continuum description of ratcheting would  require  additional continuum variables taking into account the vorticity and  the gearing between the contacts. As in the case of the shear band formation, the Cosserat theory may be a good alternative \cite{vardoulakis89}

\section{Material response to cyclic loading}
\label{material}

\begin{figure} [ht]
\begin{center}
    \psfig{file=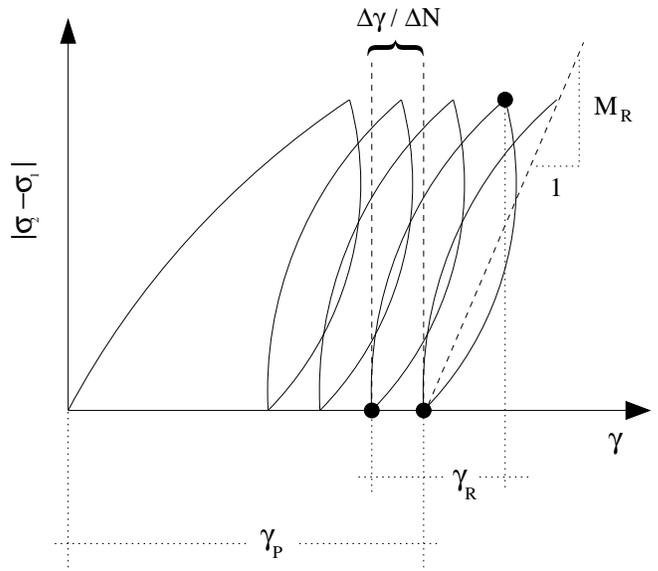,width=1.0\linewidth}
    \caption{Sketch of the typical material reaction to cyclic loading in 
the granular ratcheting. After a post-compaction stage, the system accumulates permanent strain, $\gamma_P$, at a constant strain-rate $\Delta \gamma / \Delta N$. The resilient modulus $M_R$ is also indicated, as defined in Eq. ($\ref{eq:MR}$).}
    \label{fig:cycle}
\end{center}
\end{figure}

The existence of an elastic region in the deformation of granular
materials implies that there is a finite region in the space of stress-states around the origin, in which the system reacts reversibly. Experiments and simulations show, however, that there is not such pure elastic behavior in a granular sample. Note, that this is not in contradiction with the existence of shakedown: A granular system may not accumulate any systematic
permanent deformation after one loading cycle, but will always dissipate some
energy because grain interactions are inherently inelastic. This is possible
thanks to the additional energy supplied to the system by the external
loading. In the particular case of our model, the system reaches a
visco-elastic shakedown. In this limit state, the system dissipates some energy in each cycle and the overall behavior is not elastic, but the stress-strain cycle is still hysteretic (see Figure $\ref{fig:cycle}$). Therefore we differentiate in cyclic loading between an elastic and a resilient deformation of the sample. The latter implying that no permanent deformation has been accumulated after one cycle, while the first also implies the total absence of hysteresis or memory effects in the response.

It has been recently shown that there is   a broad range of values of $\Delta \sigma$ for which a granular packing reacts to the imposed cyclic 
excitations by slowly deforming in a ratcheting regime 
\cite{alonso04,garcia-rojo04}. 
This is a quasi-periodic state, macroscopically characterized by a constant 
strain rate and a conservation 
of the shape of the stress-strain cycle (see Figure $\ref{fig:shear}$). At 
the beginning of the loading process, the system suffers a re-arrangement of 
the sliding contacts, after which they start to behave periodically within 
the loading cycles. This {\it post-compaction} process is associated to a relaxation of the strain rate and also of the dissipated energy per cycle towards a 
constant value \cite{garcia-rojo04}. This stationary value of the strain 
rate fully determines the macroscopic plastic response of the system in the ratcheting regime. At any stage of the experiment, the strain can therefore be decomposed in two well differentiated components. The irreversible plastic strain accumulated after the 
end of the current cycle  $\gamma_P$, and the recoverable resilient strain, $\gamma_R$, accumulated along the cycle. In the ratcheting regime, the strain rate ($\Delta \gamma/ \Delta N$) is approximately constant, while the latter deformation is well characterized by the resilient parameters: resilient modulus, $M_R$ and the Poisson ratio $\zeta$. The first parameter, as it appears in Figure $\ref{fig:cycle}$, is the ratio of the maximum deviatoric stress and the corresponding deviatoric resilient strain:
\begin{equation}
M_R=\frac{\Delta \sigma}{\gamma_R},
\label{eq:MR}
\end{equation}
and quantifies the overall stiffness of the material. The Poisson ratio, correspondingly, is the ratio of the horizontal ($\epsilon^R_1$) and axial ($\epsilon^R_2$) resilient strains:
\begin{equation}
\zeta=- \frac{\epsilon^R_1}{\epsilon^R_2}.
\end{equation}
It is a measure for the isotropy of the deformation. The definition of 
$\epsilon^R_1$ and $\epsilon^R_2$ is similar to that in Eq.
($\ref{eq:defstrain}$). They are both measured at the final stage of the 
loading, just before unloading starts. Similarly to Eq. ($\ref{eq:gamma}$), the resilient deviatoric strain is defined in terms of the resilient strains as 
$\gamma^R=\epsilon_2^R-\epsilon_1^R$.

\begin{figure} [ht]
  \begin{tabular}{cc}
    \psfig{file=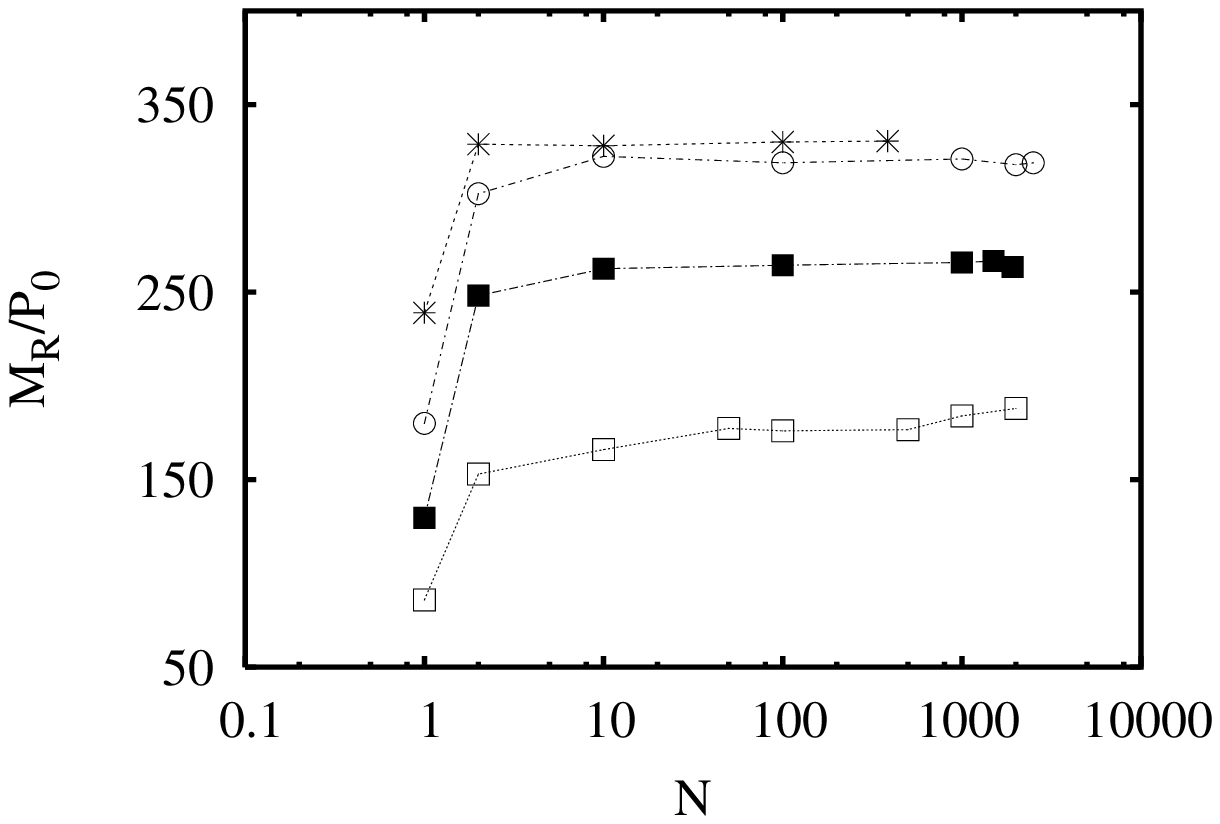,width=\linewidth,angle=0}\\
    \psfig{file=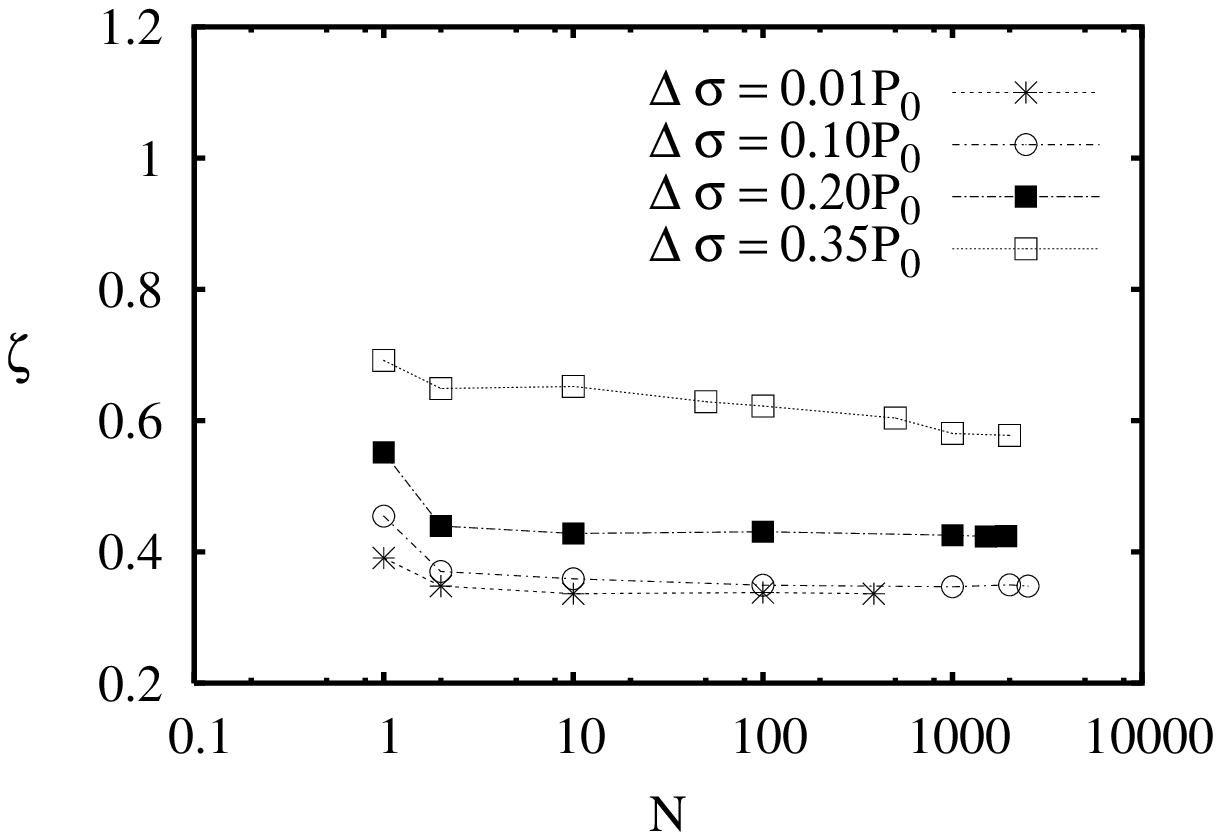,width=\linewidth,angle=0}
\end{tabular}
    \caption{Evolution of the resilient parameters with the number of 
cycles: Resilient modulus $M_R$ (top) and Poisson ratio $\zeta$ (bottom). The curves show the measures of these magnitudes for different values of the deviatoric stress $\Delta \sigma$. The data in the figure correspond to the simulation of a system with friction coefficient $\mu=0.1$, normal stiffness $k_n=2 \cdot 10^6 N/m$, normal damping $1/\gamma_n=4 \cdot 10^2 t_s$, and tangential damping $1/\gamma_t=8 \cdot 10^1 t_s$. The confining pressure is $P_0= 6 \cdot 10^{-4} k_n$.}
    \label{fig:N}
\end{figure}

As a consequence of the quasi-static change of the stresses, all the relevant time dependence occurs in the system through the number of cycles $N$. Figure $\ref{fig:N}$ shows the evolution of the resilient parameters from 
the simulations for different deviatoric stresses. For low excitations, the 
curves have already reached a {\it plateau} after a couple of cycles, 
implying that the values of $\zeta$ and $M_R$ do not apparently change as 
the number of cycles increases. In the initial post-compaction stage, the 
system accumulates more deviatoric strain in the horizontal direction (perpendicular to the direction on which the cyclic load is applied), than it does in the final stage. This explains why Poisson ratio decreases slightly in the first cycles. The resilient modulus increases, however, implying a higher stiffness of the system after the post-compaction. Although the dependence of the final values on the imposed loading will be discussed in a latter section of this paper, it should now be remarked that the number of cycles needed for the system to reach a steady resilient response increases as the imposed deviatoric stress is increased. This is clearly observed in case $\Delta \sigma =0.35$ of the figure, where even after $N=1000$ cycles, neither $\zeta$ nor $M_R$ have reached a stationary value.

The peculiar behavior of the system in the ratcheting regime allows for the 
characterization of the deformation state of the system through the strain rate and the resilient parameters. It is therefore crucial to know the influence of the confining pressure and the deviatoric stress on these parameters. For a complete review on the macroscopic factors affecting the resilient response of a granular material and some of the models proposed to account for it, we recommend references \cite{lekarpb00} and \cite{taciroglu02}.

To our knowledge, no systematic study has been carried out up to now elucidating the effect of the microscopic parameters of the system on the material reaction to cyclic loading, although they play an important role \cite{combe02, combe00}. Combe et al. have identified contact stiffness and friction as the relevant microscopic parameters in this limit. Inter-granular friction, in particular, appears then to be the dominating dissipative mechanism. The influence of contact stiffness and friction on the plastic behavior of a granular packing undergoing ratcheting will be also investigated in the following sections.

\section{Hysteretical behavior}
\label{hysteresis}
\begin{figure} [ht]
    \psfig{file=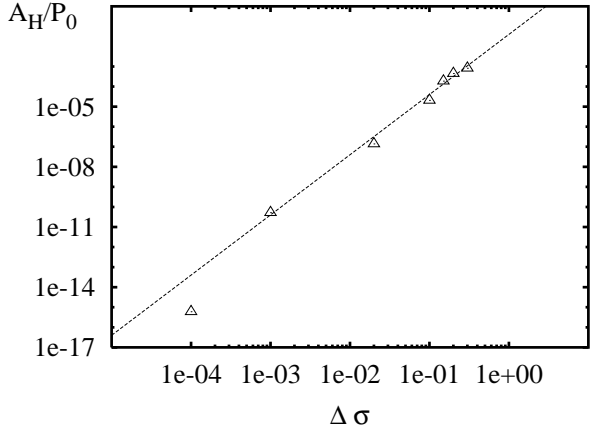,width=1.0\linewidth,angle=0}
    \caption{Variation of the area enclosed by the stress-strain cycle $A_H$, for different values of $\Delta \sigma$. The area is scaled with the confining pressure. The dashed line shows the power law $y \propto x^{3}$. The data in the figure correspond to the simulation of a system with friction coefficient $\mu=0.1$, normal stiffness $k_n=1.6 \cdot10^6 N/m$, tangential stiffness $k_t=0.33 k_n$, and normal damping $1/\gamma_n=4 \cdot 10^3 t_s$. The confining pressure is $P_0=6 \cdot 10^{-3} k_n$ and the damping coefficient $\gamma_t=8 t_s$. }
    \label{fig:energy}
\end{figure}

History dependence is one of the most essential features of granular soils. In our simple model, we have shown the existence of hysteresis both in the shakedown and in the ratcheting regime. This has forced us to identify two different components to the total strain, namely the permanent and the resilient strain. In any stress cycle, the sliding contacts behave differently in the loading and un-loading phase, leading to a different stiffness of the material in each of these phases. In this section, we are interested in the shape of the cycles and, more specifically, 
in its relationship with the evolution of the area closed by the  strain-stress loop. If we assumed that the deformation in both spatial directions is approximately the same, this area is the dissipated energy within the cycle. This energy relaxes during the {\it post compaction} from an initial high value to a constant value \cite{garcia-rojo04}, reflecting the similarity of the hysteresis loops in the ratcheting regime (see Figure ~\ref{fig:shear}). This final value is plotted in Figure ~\ref{fig:energy} for different deviatoric stress. A clear power law behavior is observed in a wide range of values above the shakedown regime.  

For the purposes that will be seen next, let us introduce the following variables:
\begin{eqnarray}
\gamma^*=\gamma_0+\frac{\gamma_R}{2}-\gamma,\\
q^*=\frac{\Delta \sigma}{2}-\frac{\sigma_2-\sigma_1}{P_0}.
\end{eqnarray}
Being $\gamma_0$ the permanent strain accumulated up to the end of the previous cycle we are interested in.

\begin{figure} [ht]
    \psfig{file=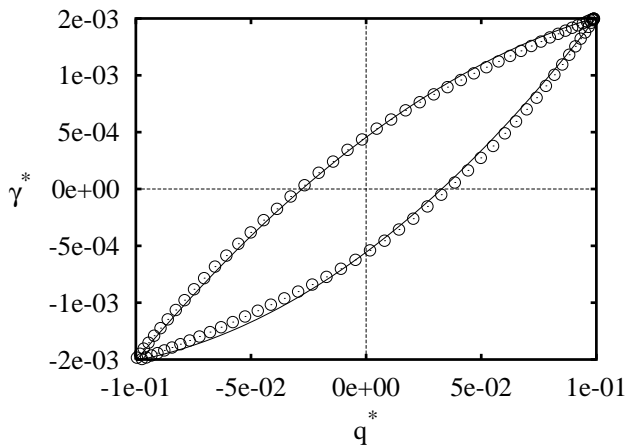,width=.7\linewidth,angle=-90}
    \caption{Hysteresis stress-strain loop in the new variables $ \gamma^*$ and $q^*$. The solid points are the result of the simulation shown in Figure ~\ref{fig:shear} ($N=1000$). The solid lines are the best-fit to the expressions (~\ref{eq:fit1}) and (~\ref{eq:fit2}). The values of the constants for the theoretical lines are $B_L= 0.04543$, and $B_U=0.05554$. }
    \label{fig:banana}
\end{figure}

We express in Figure ~\ref{fig:banana} the limit cycle on Figure ~\ref{fig:shear} on these new variables.  The best-fit curve to the points in the loading and unloading are also included. These curves can be expressed, using the scaled variables:
\begin{equation}
\gamma^*_L = \frac{1}{M_R} q^* + B_L \left ( \left (\frac{\Delta \sigma}{2} \right )^2 - {q^*}^2 \right ),
\label{eq:fit2}
\end{equation}
in the loading. And
\begin{equation}
\gamma^*_U = \frac{1}{M_R} q^* - B_U \left ( \left (\frac{\Delta \sigma}{2} \right ) ^2 - {q^*}^2 \right ),
\label{eq:fit1}
\end{equation}
in the unloading phase. $B_L$ and $B_U$ are positive constants dependent on the confining pressure, but independent on the maximum deviatoric stress ($\Delta \sigma$). Note the use of the resilient parameter $M_R$ in the previous expressions. From these formulas, it is then trivial to find the area of the cycle ($A_H$):
\begin{eqnarray}
A_H= \oint \frac{\sigma_2-\sigma_1}{P_0} d\gamma = \oint \gamma^* dq^*  ~~~~~~~~~~~~~~~~~~~ \\ = \int_{-\Delta \sigma / 2}^{\Delta \sigma /2} \left ( \gamma^*_L - \gamma^*_U \right) d q^* = ~~~~~~~~~~~~~~~~~~~~\nonumber \\
 (B_L+B_U) \left [ \left ( \frac{\Delta \sigma}{2} \right ) ^2 - \frac{\Delta \sigma}{3} \right ]_{-\Delta \sigma /2}^{\Delta \sigma /2} = \frac{ 5 (B_L+B_U)}{24} \Delta \sigma^3. \nonumber
\end{eqnarray}

Due to our definition of $q^*$ and $\gamma^*$, the area $A_H$ is in fact the same as the area enclosed by the stress-strain cycle in Figure \ref{fig:shear}. Our simple calculation explains why, given the nature of the stress-strain cycles obtained in our model, the power law behavior on Figure ~\ref{fig:energy} should be expected. The explanation shown here somehow resembles the Rayleigh law for magnetization of ferromagnetic materials under low inductions
\cite{zapperi02}. Also in this case, the hysteresis energy loss
(the area of the induction versus magnetization loop) behaves like  
the cube of the induction. This power law in ferromagnetic materials
results from the quadratic dependence of the magnetic field on the 
magnetization. This is analogous to Eqs. (~\ref{eq:fit1}) and (~\ref{eq:fit2}) except for the fact that $B_L \ne B_U $, which reflects the asymmetry of the 
loops in the granular ratcheting regime. It is interesting to observe
that the power law is identical to the one found for the dependence of
the strain-rate on the deviatoric strain, as shown in the previous
section. In fact, the closed-loop approximation given by 
Eqs. (~\ref{eq:fit1}) and (~\ref{eq:fit2}) is not strictly valid in the limit $q^* \rightarrow 0$. The error of this quadratic approximation is of the order of $\mathcal{O} ( \sigma^3) $, and must be related to the cubic dependence of the strain accumulation on the load amplitude. A micro-mechanical explanation of this Rayleigh-like law in granular ratcheting is still an open issue. 

In the ratcheting regime the factors follow $B_U > B_L$. It is still to be determined which precise effect  has the behavior of the sliding contacts on this observation. A better understanding of the nature of these constants and their dependencies on the model parameters will help to gain insight into the overall plastic response of the material.

\section{Permanent strain accumulation}
\label{plastic}

The influence of macro-mechanical magnitudes and the microscopic parameters of the model on the accumulation of permanent strain  will be shown in this section. This will be done by measuring the strain rate in simulations were the confining pressure, the deviatoric stress, the friction coefficient or the stiffness of the contacts are changed, while the rest of the parameters are kept fixed.

\subsection{Influence of the confining pressure and deviatoric stress}
\begin{figure} [ht]
\begin{tabular}{cc}
    \psfig{file=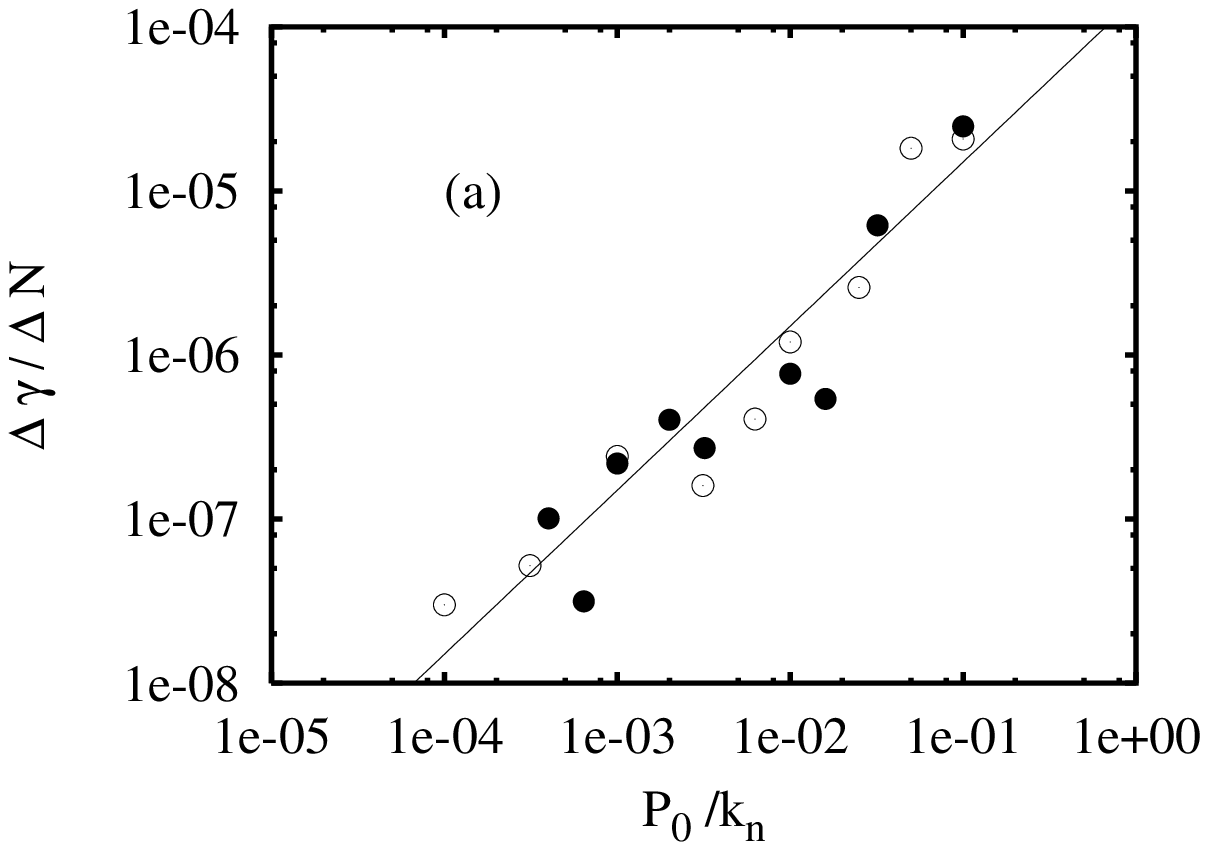,width=1.0\linewidth} \\
  \psfig{file=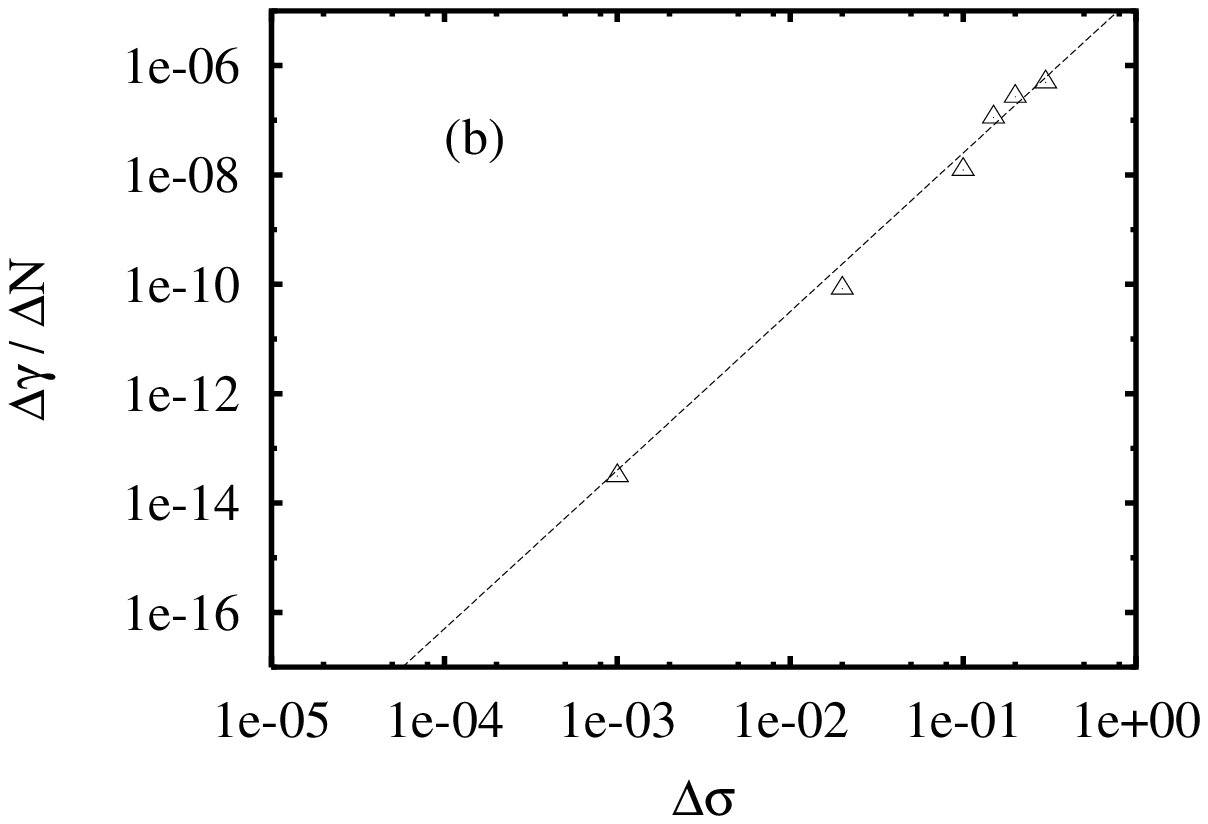,width=1.0\linewidth}
\end{tabular}
    \caption{Strain-rate dependence on the confining pressure, $P_0$, and the
    deviatoric stress $\Delta \sigma$. The solid line represents the best-fit
    power law. The simulation details are those of Figure ~\ref{fig:energy}.  Data on the top graph correspond to $\Delta
    \sigma=0.2$ and tangential damping $1/\gamma_t=8 \cdot 10^2 t_s$. Solid
    circles were obtained keeping $k_n$ constant and varying $\hat{P_0}$. The
    open circles, on the contrary, are the result of a series of simulations
    in which $k_n$ was changed. The solid line on this graph shows a linear behavior. Data for the plot on the bottom correspond to $P_0=6 \cdot
    10^{-3} k_n$ and $\gamma_t=8 t_s$. The solid line represents the power
    law $y \propto x^{3}$. This is close to the power law fitting in
    polygonal packing, whose exponent lies between $2.7$ and $2.9$ \cite{alonso04}.}
\label{fig:p0-s}
\end{figure}

Among all the possible parameters affecting the plastic behavior of a granular sample, the dependence on the confining pressure and on the deviatoric 
stress are known to be the most relevant ones \cite{lekarp00}. Since $P_0$ is measured in units of the normal stiffness, $P_0= \hat{P_0} k_n$, in our simple model, there are two equivalent ways of studying the effect of the confining pressure: On the one hand, the normal stiffness of the contact can be changed while maintaining the ratio $k_t/k_n$ constant. On the other hand, the effective pressure $\hat{P_0}$ can be increased. In order to investigate the importance of the stress history of the sample, both methods have been used and the results are shown on Figure $\ref{fig:p0-s}$.(a). In each of the simulations, the system was first homogeneously compressed, and then subjected to cyclic loading.  A power law relating the change of strain per cycle, $\Delta \gamma / \Delta N$, to $P_0/k_n$ in a wide range of values is found in our simulations. The best fit of the points leads to the linear behavior:
\begin{equation}
\frac{\Delta \gamma}{\Delta N} \propto \frac{P_0}{k_n}.
\label{eq:emplaw}
\end{equation}

Dispersion of the data with respect to the empirical law in Eq. (~\ref{eq:emplaw}), is a direct consequence of the dependence of the final strain rate on the preparation of the material. Different confining pressures imply a different post-compaction process \cite{garcia-rojo04} and therefore a different density of the sample before cyclic loading. The range of densities involved in Figure $\ref{fig:p0-s}$.(a) goes from solid fractions $\Phi = 0.82$ to $\Phi=0.9$. Our results show, in fact, that the strain-rate seems to be much more sensitive to changes in the density than the resilient parameters. This makes the investigation of the strain accumulation more difficult, limiting also the accuracy of our results on the relationship between the basic parameters of the system and the strain-rate.

This history dependence of the material is not observed in part (b) of
Figure $\ref{fig:p0-s}$, where the strain-rate accumulation is plotted versus
the deviatoric stress for the same initial configuration of disks with solid
fraction  $\Phi=0.85$. The measures indicate a clear potential dependence of
the strain-rate with $\Delta \sigma$. Also an exponential behavior (with exponent $m=2.8 \pm 0.1$) has been reported in a polygonal packing \cite{alonso04}.

\subsection{Influence of the micro-mechanical parameters}
\begin{figure}
  \begin{center}
    \psfig{file=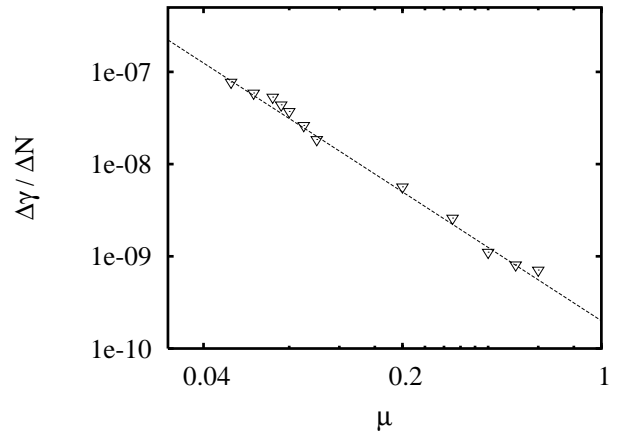,width=1.0\linewidth,angle=0}
    \caption{Dependence of the strain rate  on the friction coefficient $\mu$. The data in the figure correspond to the simulation of a system normal stiffness $k_n=1.6 \cdot 10^6 N/m$, tangential stiffness $k_t=0.33 k_n$,  normal damping $1/\gamma_n=4 \cdot 10^3 t_s$, and tangential damping $1/\gamma_t=8 t_s$. The stress conditions are $P_0= 10^{-3} \cdot k_n$ and $\Delta \sigma =0.1$. The solid fraction of the initial condition is $\Phi=0.93$. The solid line shows the law $y = x^{-2}$.}
    \label{fig:friction-s}
\end{center}
\end{figure}

The strain-rate behavior as friction changes is slightly more
complicated, if compared to the other parameters studied. For very low friction, no ratcheting is observed in the sample. Above a certain value of $\mu$,
however,  a systematic ratcheting effect can be found. For the parameters used in the simulation shown in Figure ~\ref{fig:friction-s}, this limit value is $\mu=0.05$.  The strain-rate is maximal at this friction, and (as observed in the figure) the strain-rate decreases from this point, as friction is increased. The explicit dependence on the friction coefficient follows the power law:

\begin{equation}
\frac{\Delta \gamma}{\Delta N} \propto ( \mu )^{-2. \pm 0.05}.
\end{equation}

\begin{figure} [ht]
  \begin{center}
         \psfig{file=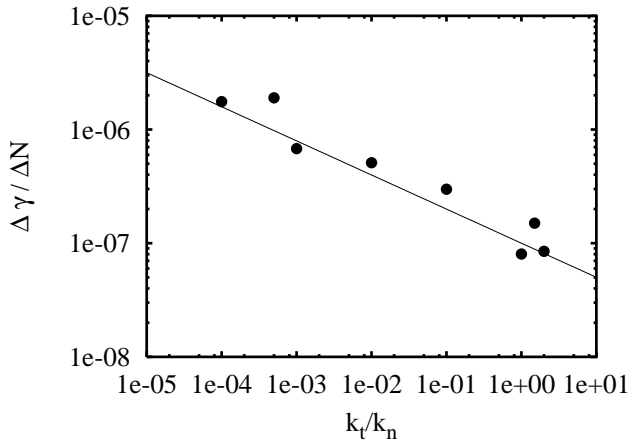,width=1.0\linewidth,angle=0} 
\end{center}
    \caption{Dependence of the strain-rate  on the stiffness ratio
    $k_t/k_n$. Data correspond to the simulation of a system with normal
    damping $1/\gamma_n=4 \cdot 10^3 t_s$, and tangential damping
    $1/\gamma_t=8 \cdot 10^2 t_s$, solid fraction $\Phi=0.845 \pm 0.005$, and
    friction coefficient $\mu=0.1$. The stress conditions are kept constant,
    $P_0=10^{-3} \cdot k_n $  and $\Delta \sigma =0.2$. The solid line represents the power law $y \propto x^{-0.3}$.} \label{fig:em-s}
\end{figure}

Figure $\ref{fig:em-s}$ shows the variation of the permanent strain 
accumulation rate with the stiffness ratio for different samples prepared 
with the same confining pressure $P_0$ and normal stiffness $k_n$. A power law
behavior with a negative exponent is found. The best fit of the points of the
figure gives:

\begin{equation}
\frac{\Delta \gamma}{\Delta N} \propto \left (\frac{k_t}{k_n} \right )^{-0.3},
\end{equation}
indicating that stronger tangential forces produce a higher rate of the deformation.

An interpretation of these power law relation could be done by exploring
the statistical distribution of the contact forces and its evolution during
the loading stage.  An important parameter is the mobilized angle $\alpha = |f_t|/f_n $, which is bounded by the sliding condition $\alpha=\mu$. The statistical distribution of this variable is rather constant except for a peak at $\mu$ given by the sliding condition. The value of this peak depends on the friction coefficient. For small values of $\mu$ a large number of contacts can reach the sliding condition so that the ratcheting response is expected to be large. For big values of $\mu$ only a few number of contacts can reach the sliding conditions, which produces a small ratcheting response. A quantitative explanation for the power law dependence will require to calculate the evolution of the  statistics of the sliding contacts and the contribution of the sliding to the global dissipation, but this is beyond the scope of this work.

\section{Resilient response}
\label{resilient}

Most theoretical models for the resilient response are based on curve fitting 
procedures, using data from biaxial or triaxial tests.  One of the most 
popular and earlier models is the so-called $k-\theta$ model \cite{hicks71}, 
in which the resilient modulus is supposed to depend only on the mean stress 
$\theta$:
\begin{equation}
M_r(\theta)=k \left ( \frac{\theta}{\eta} \right)^n,
\label{eq:kteta}
\end{equation}
where $k$ and $n$ are material constants, $\eta$ is a universal constant in
units of stress (included for normalization), and $\theta$ is the absolute
value of the first invariant of the stress tensor:

\begin{equation}
\theta \equiv | tr( \hat{\sigma})|.
\end{equation}
Many alternatives to and modifications of this model have been introduced, which are extensively used in practice \cite{hjelmstad00,nataatmadja01,lekarpb00}. One of the main restrictions of the $k-\theta$ model 
is the assumption of a constant Poisson ratio. Several studies have shown 
that the Poisson ratio is not a constant in the granular case, but varies with the applied stresses \cite{allen74}. Another drawback of the model is that the effect of the deviatoric stresses on the resilient modulus is neglected. 
A straightforward modification of the  $k-\theta$ model accounting for this latter restriction reads \cite{uzan85}:
\begin{equation}
M_r(\theta,\Delta \sigma)=k \left ( \frac{\theta}{\eta} \right)^n \left ( 
\frac{\Delta \sigma}{\eta} \right)^m.
\label{eq:kteta2}
\end{equation}
Note that, with respect to equation ($\ref{eq:kteta}$), a new material
constant $m$ has been introduced. In the simplest approximation both exponents
are assumed identical $n \equiv m$ \cite{tam88}.

The validity of the $k-\theta$ model will be checked in this section.  Note
that, in the case of cyclic loading, given a fixed $\Delta \sigma$, the
dependence of the resilient modulus on $\theta$ is similar to its dependence
on $P_0$. Results will be shown on the influence of the confining stress and 
deviatoric stress on the resilient modulus and Poisson ratio. In the latter case, it will be particularly interesting to investigate the limit of
validity of the common assumption of a constant 
Poisson ratio for granular matter.

\subsection{Influence of the confining pressure}
\begin{figure} [ht]
  \begin{tabular}{cc}
    \psfig{file=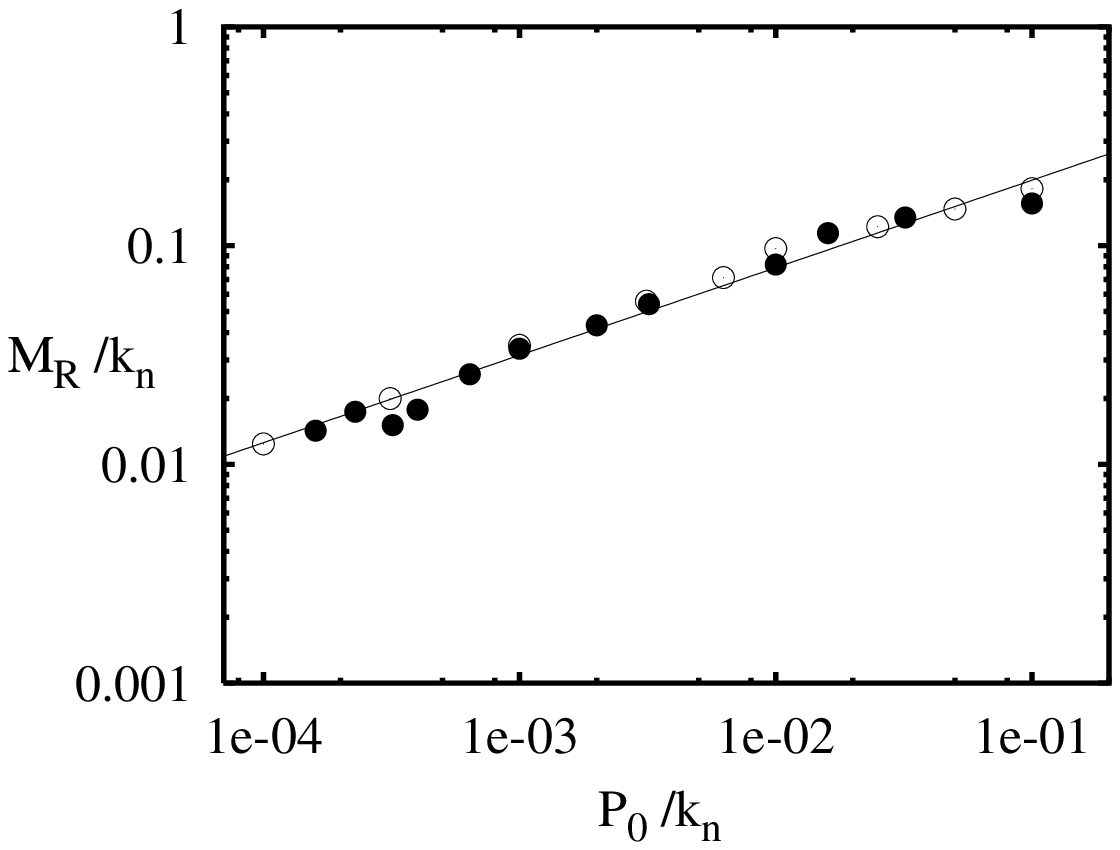,width=1.0\linewidth,angle=0}\\
    \psfig{file=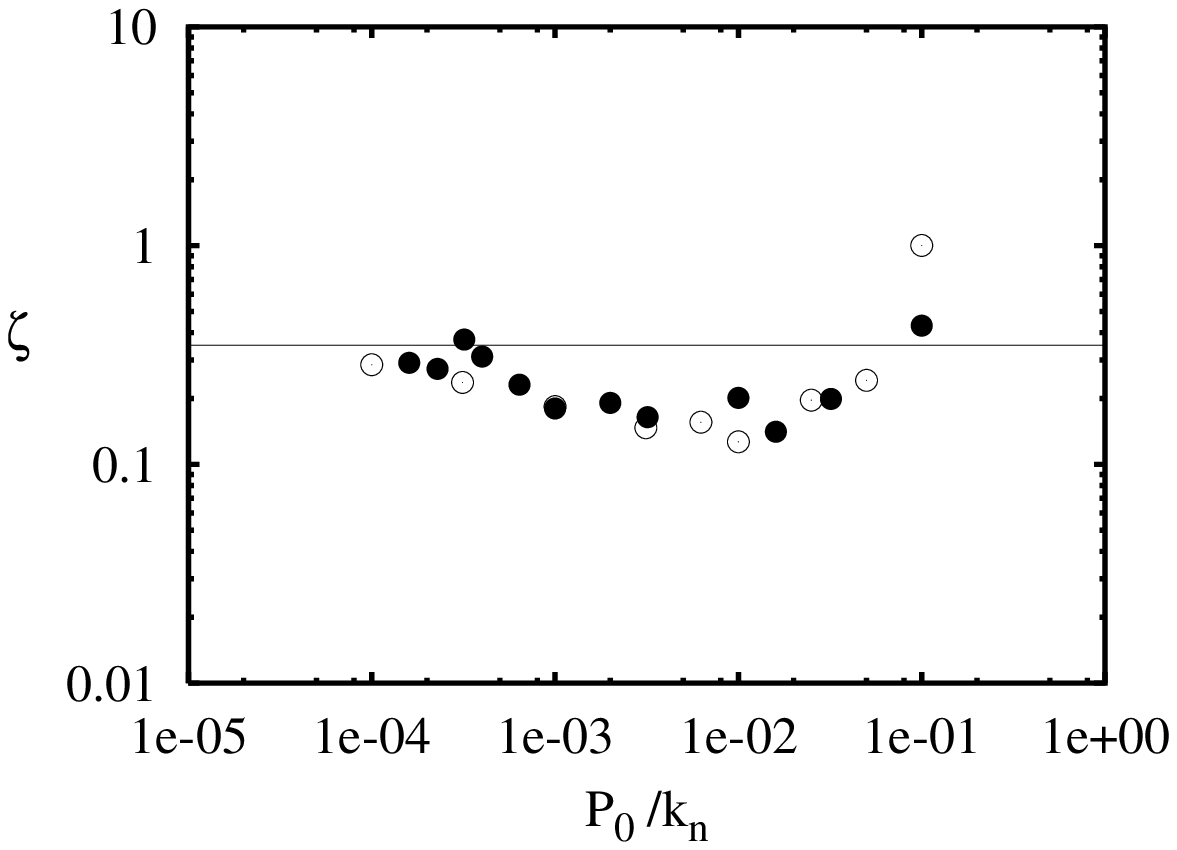,width=1.0\linewidth,angle=0}
\end{tabular}
    \caption{Variation of the resilient parameters with the confining 
pressure $P_0$: resilient modulus $M_R$ (top) and Poisson ratio $\zeta$ (bottom). The conditions of the simulation are the same as in Fig. $\ref{fig:p0-s}$. The line in the left plot is the best fit to the $k-\theta$ model. The 
solid line in the right figure is the value $\zeta=0.35$, estimation for the
Poisson ratio of granular materials. The different symbols refer to two
different methods explained in the text to study the influence of the
confining pressure on the system.}
    \label{fig:p0}
\end{figure}

Figure $\ref{fig:p0}$ indicates that the $k-\theta$ model is in fact a very good approximation in the ratcheting regime for a wide range of pressures of
$P_0$. The best fit to the empirical law of Eq. (~\ref{eq:kteta}), gives
$n=0.34 \pm 0.02$. This value agrees well with the experimental values in
\cite{allen74}, where results on gravel show a power law with exponent
$n=0.31$.

The Poisson ratio behaves in a completely different way. For low pressures, it
decreases gradually as the pressure becomes higher. For $P_0 > 0.01 k_n$,
however, there is a change on the trend, and  $\zeta$  grows fast with $P_0$. This reflects a higher anisotropy of the deviatoric strain in systems compressed
under a high pressure. Nevertheless, our results justify the use of a constant
value of $\zeta$ in a first approximation, for a wide range of $P_0$, $10^{-4}
k_n < P_0 < 10^{-2} k_n$. The most common estimate ($\zeta=0.35$), however, slightly overestimates the values obtained in most of our simulations.

\subsection{Influence of the deviatoric stress}
\begin{figure} [ht]
  \begin{tabular}{cc}
    \psfig{file=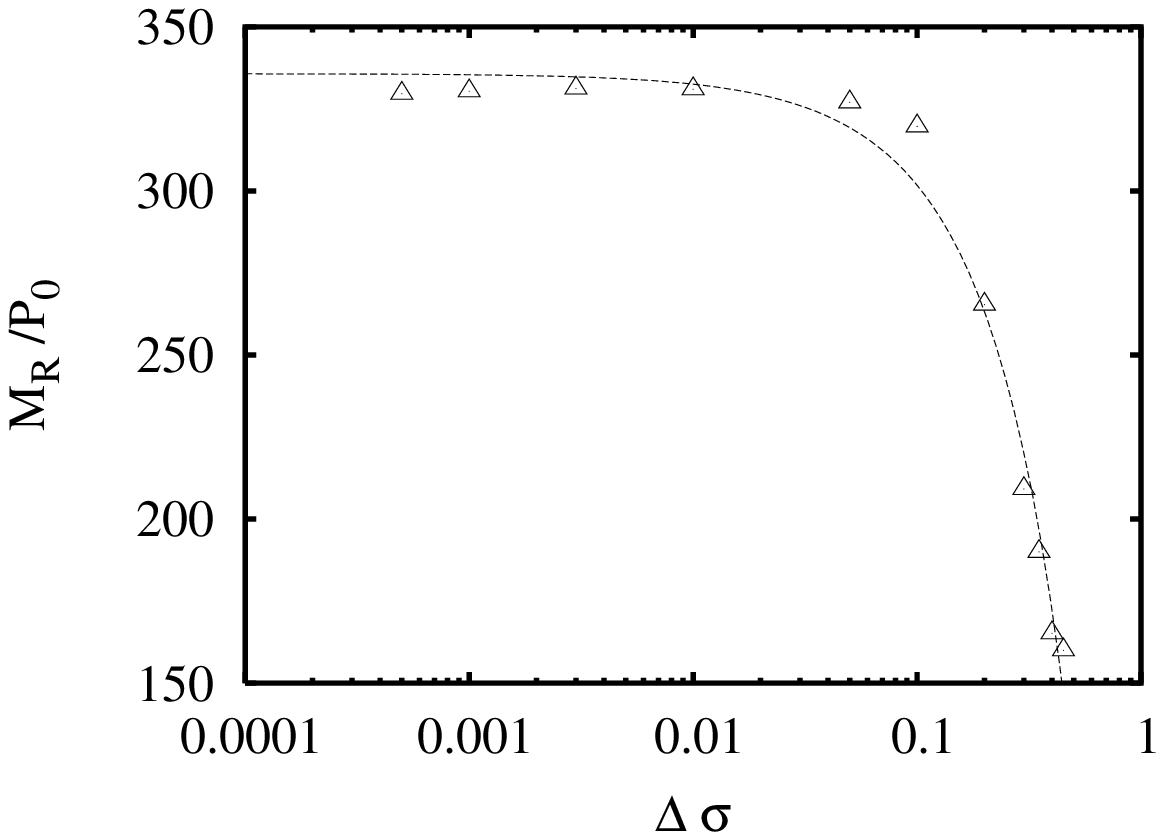,width=1.0\linewidth,angle=0}\\
    \psfig{file=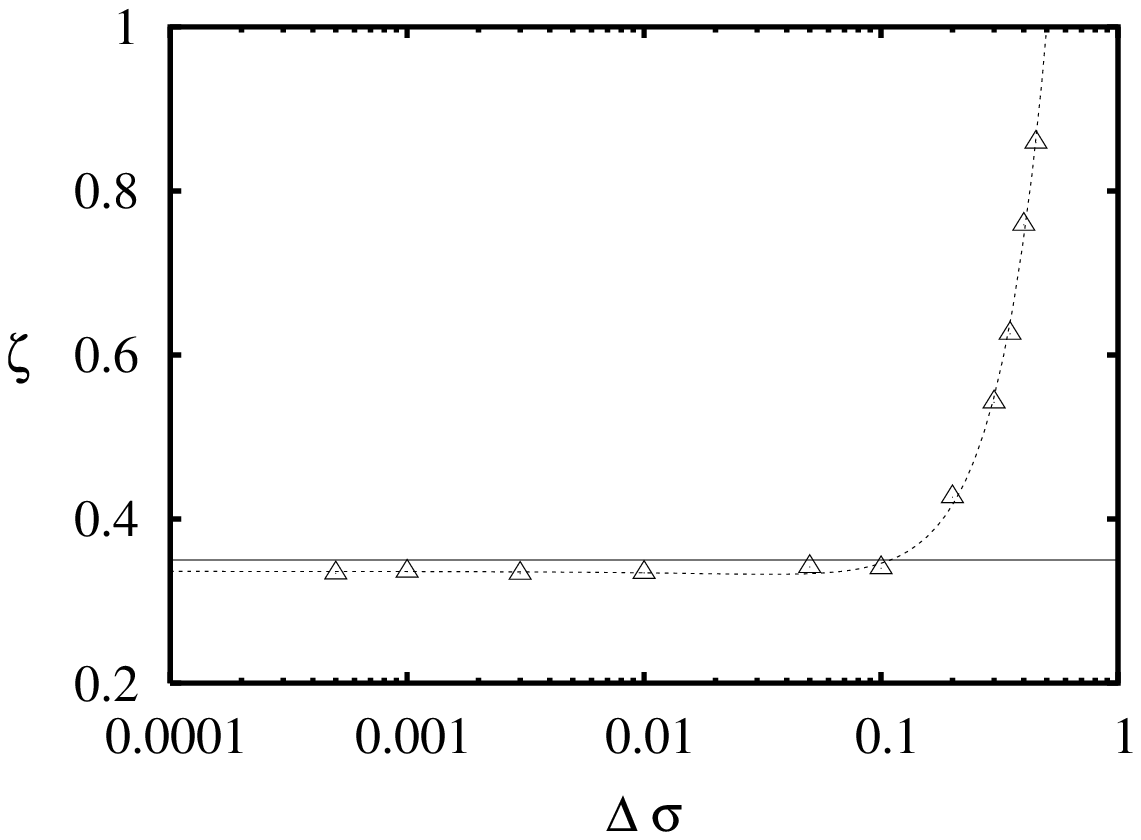,width=1.0\linewidth,angle=0}
\end{tabular}
    \caption{Variation of the resilient parameters with the loading intensity
    $\Delta \sigma$. The simulation details are similar to those in Figure
    ~\ref{fig:p0-s} but with $k_n=2 \cdot 10^6 N/m$, and $P_0=6 \cdot 10^{-4}k_n$.  The
    best fit curve to a second order polynomial is plotted for the values of
    $M_R$ in the top graph. In the bottom (Poisson ratio),
    the solid line corresponds to the value $\zeta=0.35$ and the dotted line
    to the best fit to equation $y(x)=a+bx+cx^2$ (details are given in the
    text).}
    \label{fig:loading}
\end{figure}
Two stages are clearly distinguished in the behavior of the resilient
parameters as a function of $\Delta \sigma$. For low values of the deviatoric
stress, close to the shakedown regime, the resilient parameters remain
approximately constant.  Poisson ratio, remains closer to the indicated
value $\zeta \approx 0.35$ which is the empirical fixed value usually assumed
for unbound granular matter \cite{allen74}. This value is shown in Figure
$\ref{fig:loading}$ with a solid line. For $\Delta \sigma > 0.1$, however,
$\zeta$ shows a strong dependence on the deviatoric stress $\Delta \sigma$.

A simple empirical polynomial law is proposed in reference \cite{allen74} for
the dependence of $\zeta$ on the ratio of the deviatoric and volumetric
stresses. Although the range of values studied in this experiment is larger
than the one presented here, our results confirm that the values of the Poisson
ratio follow a second order polynomial law on $\Delta \sigma$, being the
best-fit curve $\zeta = 0.336 (\pm 0.001)- 0.208 (\pm 0.001)  \Delta \sigma +
3.061 (\pm 0.001) (\Delta \sigma)^2$. This curve is plotted in the lower part
of Figure $\ref{fig:loading}$.

As opposed to the behavior of Poisson ratio, the resilient modulus decreases
as $\Delta \sigma$ increases. the dependence is also polynomial. In
Figure ~\ref{fig:loading} (top), the curve $y(x)= 335.7  -316.8 x + 229.1 x^2$ is plotted.  Note that this result disagrees with the simplification of the
generalized $k-\Theta$ model ($m \equiv n$) of equation
($\ref{eq:kteta2}$). The general law seems to be a better approximation in a
wide range of values of the deviatoric stress, where the system shows neither
collapse nor shakedown.

The dependence of the resilient parameters on the deviatoric 
stress results from the anisotropy induced in the contact network 
for large deviatoric loads. Near failure, a significant number 
of contacts are open in the perpendicular direction of the load, resulting 
in a decrease of the stiffness as shown in the top of Figure ~\ref{fig:loading}. The increase of the Poisson ratio in the bottom of this figure is 
consequence of the formation of force chains, which enhance the anisotropy 
and leads to an increase of the effective Poisson ratio. A detailed 
description of the effect of these force chains in the resilient response 
would require a detailed evaluation of the relation between the anisotropy of the contact network and the parameters of the anisotropic elasticity via fabric tensors \cite{luding04,alonso04c}.

\begin{figure} [ht]
  \begin{tabular}{c}
    \psfig{file=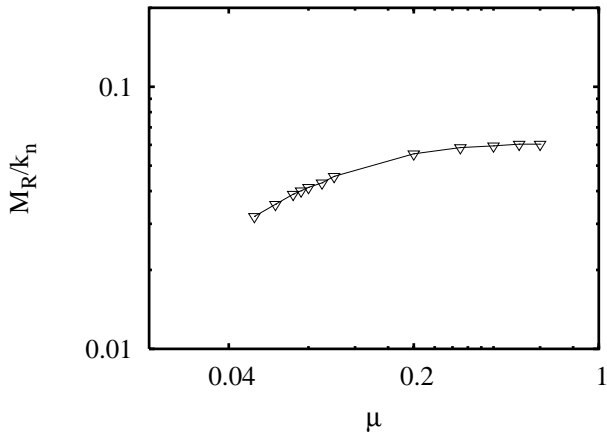,width=1.0\linewidth,angle=0}
\end{tabular}
    \caption{Variation of the resilient modulus with the static friction 
coefficient $\mu$. The conditions of the simulation are the same as in Fig. $\ref{fig:friction-s}$.}
    \label{fig:friction}
\end{figure}

\subsection{Influence of the micro-mechanical parameters}

Figure $\ref{fig:friction}$ shows the change of the resilient modulus with friction. $M_R$ grows for small frictions. However, the curve seems to reach a saturation level for frictions $\mu \approx 0.4$.

\begin{figure} [ht]
  \begin{tabular}{cc}
    \psfig{file=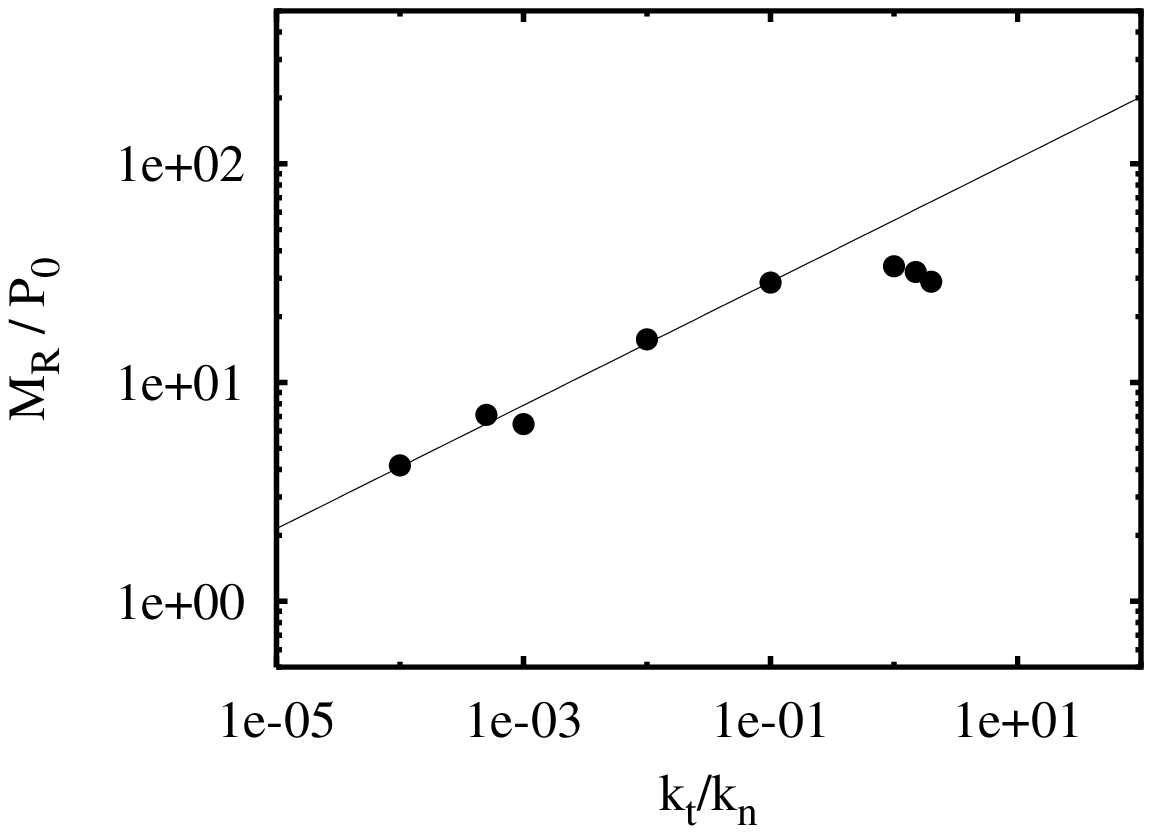,width=1.0\linewidth,angle=0}\\
    \psfig{file=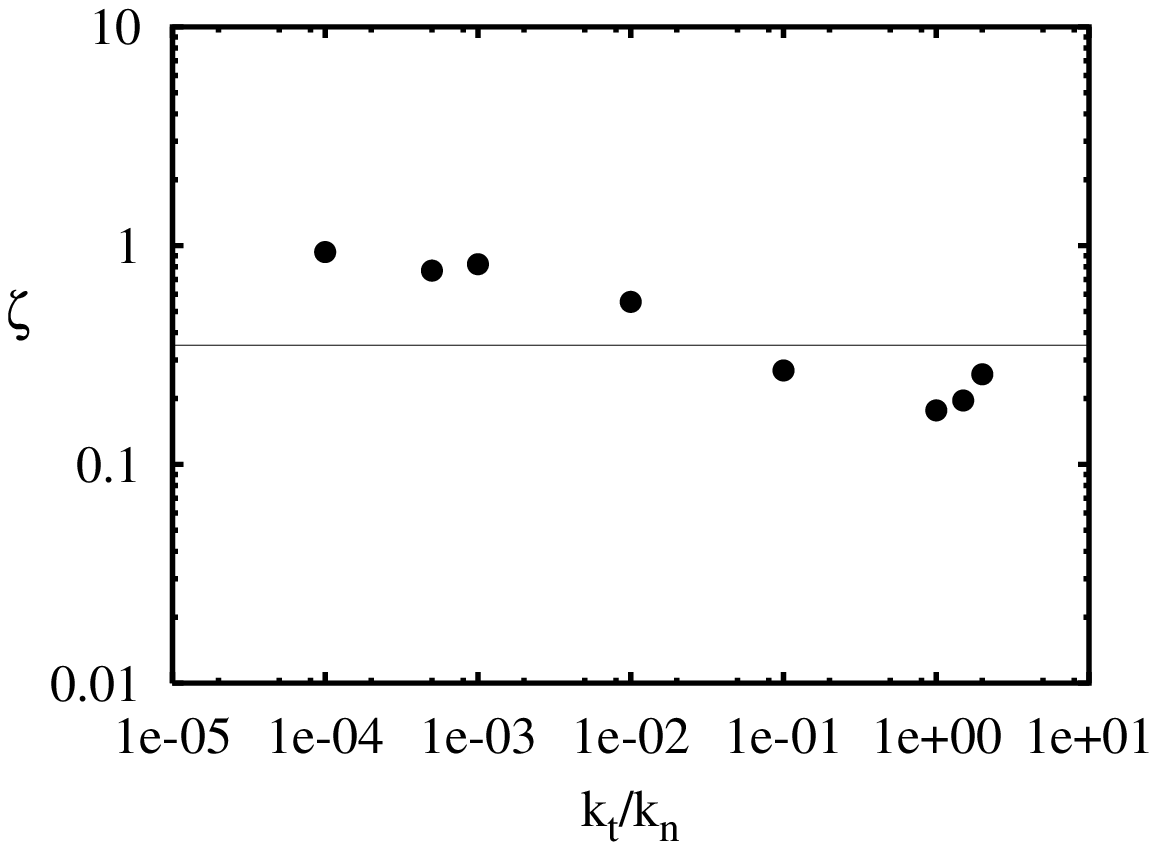,width=1.0\linewidth,angle=0}
\end{tabular}
    \caption{Influence of the  ratio of contact stiffness $k_t/k_n$ on the resilient parameters. The details of the simulation are those of Fig. $\ref{fig:em-s}$. The solid line shows a power law with exponent $0.28$ in the top. The one at the bottom marks the value $\zeta = 0.35$.}
    \label{fig:knkt}
\end{figure}

Changing the ratio of contact stiffness (Fig.~\ref{fig:knkt}), a power law dependence of $M_R$ is observed for $k_n/k_t < 0.1$,  $M_R \propto \left ( \frac{k_n}{k_t} \right )^{0.28}$, being the exponent $0.28 \pm 0.03$. For stiffness ratios closer to unity $k_t/k_n \approx 1$, the resilient modulus remains approximately constant or even decreases. The Poisson ratio also appears to be constant for $k_t < 10^{-3} \cdot k_n$. Above  $k_t/k_n= 0.001$, $\zeta$ decreases to values below the reference value $\zeta=0.35$. For $k_t \ge k_n$, $\zeta$ starts growing again.

\section{Discussion and final remarks}
\label{discussion}

A characterization of the material response in the 
granular ratcheting has been presented in terms of the strain-rate, resilient modulus and the Poisson ratio. Studying the dependence of these parameters on the conditions of the biaxial test (stress configuration) and the main 
microscopical constants of the sample (friction and contact stiffness) we 
confirmed the persistence of the granular ratcheting in many 
different conditions and systems.

Given a compressed sample subjected to a biaxial test in which a
cyclic loading is switched on, the system adapts to the new situation accumulating deformation and dissipating energy at a relatively high 
rate. After this {\it post-compaction} stage, the dissipated energy, both 
resilient moduli and the strain-rate reach stationary 
values. The duration of the adaptation stage  basically depends on the 
deviatoric stress, and is usually shorter for the resilient moduli than 
for the strain-rate \cite{garcia-rojo04}. If the deviatoric stress is small 
enough, the perturbation introduced by the cyclic loading shakes down. The 
material adapts to the new situation so that there is no 
further accumulation of permanent strain. Above this limit the material accumulates a certain amount of strain in each cycle. If the stress is below the collapse limit, the permanent strain accumulated after each cycle is constant. This is the so-called granular ratcheting, which has been described both experimentally \cite{werkmeister01,werkmeister04} and in simulations \cite{alonso04,garcia-rojo04}. 

Identical repetition of the strain-stress cycles is among the main 
characteristics of the granular ratcheting. This periodicity reflects the 
weak dependency of the resilient moduli on the stress history and, in the 
particular case of cyclic loading, on the number of applied cycles 
\cite{lekarp00}. In all the simulations, a steady and stable resilient 
response is reached after some initial cycles. This kind of simple behavior is 
expected as long as the applied deviatoric stress remains below the collapse 
limit. Although many factors may influence the plastic response of the 
system, there is a simple characterization of the deformation in the 
ratcheting regime, in terms of the strain-rate and the resilient moduli. 
This description takes advantage of the empirical fact that these magnitudes do not change in the ratcheting regime. We have investigated both micro-mechanical and macro-mechanical factors influencing the plastic response of the material, i.e. the dependency on the number of cycles, static friction, the confining pressure, the deviatoric stress and the stiffness.

It was shown that the use of a constant Poisson ratio is a good 
approximation in most cases.  It seems to be unsuitable, however, for very high confining pressures, very high deviatoric stresses, or for low values of the friction coefficient. The value for $\zeta$ estimated through our simulations would be slightly below the empirical value $0.35$, assumed in many models of the resilient response of granular materials. This might be a consequence of the simplicity of the visco-elastic model, which does not include all the mechanisms involved in a real biaxial experiment.

$M_R$ is a measure of the macroscopical stiffness of the material. Our results show that it is higher for strongly frictional materials. We also found that although preparing the sample with a higher confining pressure increases its stiffness, increasing the deviatoric stress reduces the stiffness of the packing.

Both the strain-rate and the resilient modulus $M_R$ show a power law 
dependence with the confining pressure and the ratio of contact stiffness. The power law is similar for both magnitudes in the case of the confining pressure, but they have an opposed dependence on $k_t/k_n$. The dependence of $M_R$ on the deviatoric stress is a second order polynomial. The generalization of the $k-\theta$ model of equation ($\ref{eq:kteta2}$) is not sufficient for our system, although Eq.(~\ref{eq:kteta}) is a good approximation in many situations.

Re-analyzing our results on the strain-rate, we can summarize them in the formal expression:
\begin{equation}
\frac{\Delta \gamma}{\Delta N} \propto \frac{P_0}{\mu^2} \left ( \Delta \sigma \right ) ^3 \left ( \frac{k_n}{k_t} \right )^{0.3}.
\end{equation}

A direct relationship has been shown between this dependence, the power law behavior of the dissipated energy per cycles as a function of the deviatoric stress imposed, and the systematic accumulation of permanent strain. Although the resilient parameters are not much affected by the stress history of the material, the strain-rate is strongly dependent on it, complicating therefore the systematic investigation of the plastic response. In this context, it would be necessary to measure in more detail the influence of density and polydispersity on the possible shakedown of the material. The history dependence of the plastic response of the system is of vital importance to technical implications. 
Future topics for investigation include the study of the shakedown-ratcheting transition as a function of the friction and the loading intensity. The influence of the system size and the dependence on the damping constants are subjects of current work.

{\bf Acknowledgments:} The authors would like to thank Proferssors Deepak Dhar and Ioannis Vardoulakis for very useful discussions.  They also want to acknowledge the EU project Degradation and Instabilities in Geomaterials with Application to Hazard Mitigation (DIGA) in the framework of the Human Potential Program, Research Training Networks (HPRN-CT-2002-00220).


\end{document}